# Exploring Stress among International College Students in China

Omogolo Omaatla Morake[1], Mengru Xue[1, *]

1. Zhejiang University Ningbo Innovation Center, China

## Abstract

Psychological stress encompasses emotional tension and pressure experienced by people, which usually arises from situations people find challenging. However, more is needed to know about the pressures faced by international college students studying in China. The goal of this study is to investigate the various stressors that international college students in China face and how they cope with stress (coping mechanisms). Twenty international students were interviewed to gather data, which was then transcribed. Thematic analysis and coding were applied to the qualitative data, revealing themes related to the causes of stress. The following themes emerge from this data: anticipatory anxiety or future stress, social and cultural challenges, financial strain, and academic pressure. These themes will help understand the various stressors international college students in China face and how they try to cope. Studying how international college students in China cope with challenges can guide the development of targeted interventions to support their mental health. Research suggests that integrating aesthetics and connectivity into design interventions can notably improve the well-being of these students. This paper presents possible future design solutions, leveraging the aesthetics of connectivity to empower students and enhance their resilience. Additionally, it aims to provide valuable insights for designers interested in creating solutions that alleviate stress and promote emotional awareness among international students.

## Introduction

Mental health represents a crucial concern for international students pursuing studies abroad, including those in China, a country that holds the third position globally in terms of higher education enrollment and has witnessed a growth in student intake over time [1, 2]. This study sets out to explore the various stressors that international college students in China face and how they cope with them. It also explores how design interventions can support their mental health and well-being.

Stress, anxiety, and mental health issues are universal concerns for international college students, regardless of their study destinations. It delves into the specific challenges these students encounter and compares them with those faced by their peers in other popular study destinations like the United States and the United Kingdom. Additionally, the study seeks to uncover the coping strategies international students employ to manage their stress. This study provides evidence that international college students in China are also facing some form of stress, like their peers in other countries like the United States of America, Russia, and Canada [3, 4]. The primary goal is to fill the knowledge gap concerning the stress levels of international college students in China, an area that has been overlooked in existing research. The research further contributes to knowledge about international students' challenges in China and provides insights for future design interventions to support their well-being.

# Methodology

Data was collected through face-to-face interviews with 20 international students from various continents, using open-ended questions to gather insights into their stressors and coping strategies. Thematic analysis was utilized to examine qualitative data to identify recurring themes.

Qualitative data was collected for this study through face-to-face interviews using open-ended questions. This method was chosen to build rapport more easily with the participants, get immediate feedback, and have a higher level of engagement which is difficult to achieve with quantitative methods. Furthermore, the open-ended nature of qualitative research allows for flexibility during data collection and analysis, enabling unexpected themes or issues that arise during interviews to be easily followed up. Numerous ethical considerations were considered for this study, such as finding an appropriate setting where participants felt most at ease, obtaining their consent to an audio recording, and emphasizing that anonymity is paramount.

# Results and Analysis

The thematic analysis revealed five main sources of stress: academic pressure, financial concerns, future worries, adaptation challenges, and social/relationship stress. Academic pressure included sudden submissions and scholarship requirements. Financial concerns stemmed from living expenses and medical bills. Adaptation challenges included language barriers and homesickness. Social stress was influenced by negative interactions with friends and classmates. This is depicted in Table 1 below with both basic and exemplary codes.

**Table 1**

Thematic Analysis of The Sources of Stress

| Basic theme(s) | Basic code(s) | Exemplar code(s) |
|---|---|---|
| Academic pressure | C1: Coping with the college education system, e.g., sudden submissions | "Sometimes you don't sleep the whole night; you are there trying to figure out how to submit that thing quickly. So, it gives a lot of stress." [P20]<br>"Yeah, actually, I'm always stressed when it is the last days and before, like submitting an assignment or a final project" [P5]" |
| | C2: Pressure to do well due to scholarship requirements | "You have that stress during the exam. You want to come up with the highest score so that you can be on the list of people who get the scholarship." [P20]<br>"So, we have the lab for all the software engineering, including computer science. So as an international student, I was not allowed to use the lab just like everyone else, so this kind of situation creates stress on you" [P20] |
| | C3: Unequal opportunities compared to Chinese students | "I could see that the Chinese students have many opportunities, opportunities than I have, and they could get better services than the international students had. Chinese students get internships easily and are more prepared to go on to another level; maybe if you have a bachelor's degree, they will prepare you for the master's degree. It was stressful at that moment |
|  |  | because maybe you are alone and need to figure out the next step alone." [P3] |
|  | C4: Difficulty in understanding everything in class due to factors like language barrier and teachers' accent | "Oh my God. I feel like my case was different from maybe other bachelor's degrees international students because I was doing my Bachelor in Chinese; I think I experienced much more stress than other international students." [P3] |
| Financial concerns | C1: Financial strain due to living expenses and medical bills | "Stress like where I see myself after five years, do I have much money that I will support my family?" [P8] |
|  | C2: Dependency on scholarship/ Not being allowed to work part-time jobs | "So, when I'm in China, they give me 1700 RMB per month. So, this is not enough for me, and I'm not allowed to work here. So, the only stress that's come to me is the financial thing." [P15] |
|  | C3: Difficulty in receiving money from family | "Finances are always a stress because even now if I have my money that my family sent me, I have difficulty receiving that money." [P20] |
| Future worry/Anticipatory anxiety | C1: Tension and worry for the uncertain future | "Stress like where I see myself after five years, do I have much money that I will support my family?" [P8] |
| Adjustment/Adaptation challenges | C1: Language barrier | "I feel somehow some stress due to some barriers and some unexpected things, like language barrier…" |
|  | C2: Homesickness | "When I came to China that time, I felt homesick because it was the first time I came so far from my home, I lived with my parents and suddenly, I was apart from them, so I was not able to do the things because of the stress and because I'm a very emotional person. So, I started crying in my room." [P12] |
|  | C3: Struggling with food in China, e.g. because of religion and differences in food | "We really struggle with food uh because in our country, we eat wheat bread, like chapatti, and people here like to eat pork uh and this kind of stuff, but in our culture, we never eat pork and this kind of things." [P8] |
|  | C4: Difficulty in public offices/services e.g. the bank | "Sometimes I have some difficulties here. Like for a bank account, I have so much stress." [P5] |
| Social / Relationship Stress | C1: Negativity from some friends and colleagues | "And when you are with your other friends, they just discuss every negative thing, they are just serious, and discussion about topics like studies is very hard this and that, I have this problem. They are always talking about the problem and in this environment, you think like oh I am also in a problem." [P19] |
|  | C2: Unfriendly classmates/environment | "I met some other fellows, they just never talk to anyone, and they just do their own thing. And when you are in that type of environment is very bad for you." [P8] |

| | C3: Having a hard time because their partner is in another country | "Family things, I feel like I'm pretty strong at those things, but about my girlfriend, I feel like I'm the most vulnerable one. I want her to be here by my side. That's all I want." [P15] |

## Coping Mechanisms

Students employed various coping mechanisms such as social media use, emotional eating, physical activities, and sleep. Social media and digital entertainment were frequently used to distract from stress. Emotional eating was common among female students, while physical activities and religious practices were also mentioned.

• Social media and Digital Entertainment: Watching movies, listening to music, and engaging with social media to distract from stress.

• Emotional Eating: Binge eating snacks and junk food, particularly among female students.

• Physical Activities: Engaging in physical activities such as sports, gym workouts, yoga, and cycling was one of the coping mechanisms identified.

• Sleeping: Taking naps or sleeping for extended periods to cope with stress. • Religious/Spiritual Activities: Engaging in prayer and watching spiritual videos.

## Design Interventions

Students suggested potential design interventions such as stress measurement tools, mood enhancers, relaxation aids, and VR spaces to feel at home. Connectivity aesthetics in design can empower students and enhance their resilience.

• Stress Measurement Tools: Wearable devices or non-mobile tools to monitor stress levels and provide feedback.

• Mood Enhancers: Interactive devices to help improve mood and relieve stress.

• Consultation and Therapy: Office spaces for students to interact with an international psychologist who can understand them better or a Chinese psychologist with global experiences.

• Interactive Platforms: Spaces (physical or online) for students to share ideas, talk, and get advice from peers.

• Relaxation Aids: Tools and spaces designed to help students relax during stressful periods.

• VR Spaces: Virtual reality environments designed to feel like home, providing comfort and reducing homesickness

## Discussion

The study identifies common stressors among international students in China and suggests targeted design solutions to alleviate stress. International students' possible designs and solutions regarding stress relief can be categorized into tangible devices, social interventions [5], space interaction, AI agents who understand their culture, and wearable devices. Everyday wearables can be used to reduce stress, as proposed by this paper [6]. According to Agnes and Marc's findings [7], although social media has its negatives, it can still offer platforms for social interventions, such as social connections and online communities, which can enhance mental health. Some research shows the potential of AI agents to reduce stress for students [8]. Exposure to different VR scenes and environments [9], and training in respiration using specially designed tangible and interactive methods [10], can also aid in stress reduction. These solutions can provide knowledge for future designers who wish to design for international students. Future designs could further explore various avenues for stress relief among international students, including digital aesthetics and games [11], biofeedback devices, designated stress-management spaces [12], or even collective stress for students [13]. The findings from this study have led to the conceptualization of design interventions based on the input from the students gathered during the interview. However, for future work, these suggested interventions will require further development and empirical testing to assess their effectiveness in relieving stress among international students. Furthermore, the research will benefit from integrating quantitative data with qualitative perceptions, providing a more comprehensive viewpoint on the stress levels encountered by international students and the effectiveness of their coping mechanisms. This dual approach will enable the study to extend its findings to a larger population, enhancing the data's generalizability.

## Conclusion

Understanding the unique stress factors of international students in China is crucial for developing effective design interventions to support their well-being. The study calls for future research to create tailored solutions leveraging technology and innovative design to support the well-being and resilience of these students, thereby fostering a supportive environment for their academic and personal success.

## Acknowledgements

This research was supported by Ningbo Innovation Center, Zhejiang University, China (Funding Number: 1140557B20220120). We would like to express great thanks to our sponsor as well as all who provided support.